\documentclass[preprint,showpacs,preprintnumbers,amsmath,amssymb]{revtex4}

\usepackage{graphicx}% Include figure files
\usepackage{dcolumn}% Align table columns on decimal point
\usepackage{bm}% bold math

\begin{document}
\def\brho{{\hbox{\boldmath $\rho$}}}
\def\bsb{{\hbox{\boldmath $\beta$}}}
\def\bsk{{\hbox{\boldmath $k$}}}
\def\bsp{{\hbox{\boldmath $p$}}}

\title{Interferometry signatures for QCD first-order phase transition
in heavy ion collisions at GSI-FAIR energies}

\author{Li-Li Yu$^1$}
\author{M. J. Efaaf$^2$}
\author{Yan-Yu Ren$^{1}$}
\author{Wei-Ning Zhang$^{1,2}$\footnote{wnzhang@dlut.edu.cn}}

\affiliation{ $^1$Department of Physics, Harbin Institute of
Technology, Harbin, Heilongjiang 150006, China\\
$^2$School of Physics and Optoelectronic Technology, Dalian
University of Technology, Dalian, Liaoning 116024, China }

\date{\today}

\begin{abstract}
Using the technique of quantum transport of the interfering pair we
examine the Hanbury-Brown-Twiss (HBT) interferometry signatures for
the particle-emitting sources of pions and kaons produced in the
heavy ion collisions at GSI-FAIR energies.  The evolution of the
sources is described by relativistic hydrodynamics with the system
equation of state of the first-order phase transition from
quark-gluon plasma (QGP) to hadronic matter.  We use quantum
probability amplitudes in a path-integral formalism to calculate the
two-particle correlation functions, where the effects of particle
decay and multiple scattering are taken into consideration.  We find
that the HBT radii of kaons are smaller than those of pions for the
same initial conditions.  Both the HBT radii of pions and kaons
increase with the system initial energy density.  The HBT lifetimes
of the pion and kaon sources are sensitive to the initial energy
density.  They are significantly prolonged when the initial energy
density is tuned to the phase boundary between the QGP and mixed
phase.  This prolongations of the HBT lifetimes of pions and kaons
may likely be observed in the heavy ion collisions with an incident
energy in the GSI-FAIR energy range.
\end{abstract}


\maketitle

\section{Introduction}

One of the central goals of high enery heavy ion collisions is to
find and quantify the QCD phase transition from hadronic matter to
quark-gluon plasma (QGP).  Under the assumption of first-order phase
transition there is a mixed phase of the QGP and hadronic gas.  In
the absence of pressure gradient, a slow-burning fireball is
expected when the initial system is at rest in the mixed phase, and
this leads to a considerable time-delay of the system evolution
\cite{Pra861,Ber89,Hun95,Ris96,Sof02,Zsc06}.  It is therefore of
interest to investigate the time-delay signatures for the
first-order phase transition.

Two-particle Hanbury-Brown-Twiss (HBT) interferometry is a useful
tool for detecting the space-time structure of particle-emitting
sources in high energy heavy ion collisions
\cite{Won94,Wie99,Wei00,Lis05}.  For the first-order phase
transition between the QGP and hadronic matter, the time-delay of
the system evolution may prolong the emission duration of particles
and lead to unusually large HBT lifetime, as compared to a crossover
transition or a hadron gas without the QCD phase transition
\cite{Pra861,Ber89,Hun95,Ris96,Sof02,Zsc06}.  It is known that the
phase transitions occurring in the heavy ion collisions at RHIC and
top SPS energies are crossover and in small baryon density regions.
Moreover, at AGS energies the systems are almost in hadronic phase
although with higher baryon densities.  The future Facility for
Antiproton and Ion Research (FAIR) at GSI with heavy ion beams from
2 -- 45$A$ GeV will provide an opportunity to explore the
first-order QCD phase transition at high baryon densities
\cite{Hoh05,Pet06,Fri06,Ars07,Ros07,Hen08}.  Investigating how the
HBT results variate with the incident energy and what the HBT
signatures the first-order phase transition is possessed of in the
GSI-FAIR energy range is thereby the subject of this work.

In high energy heavy ion collisions the final particles include the
contributions of direct production and excited-state particle decay.
Also, the particles will subject to the multiple scattering with
other particles when they propagate in the system up to the thermal
freeze-out.  In Ref. \cite{Yul08} a HBT analysis technique with
quantum transport of the interfering pair was developed to
investigate the effects of particle decay and multiple scattering on
the extracted HBT radii in the heavy ion collisions at AGS and RHIC
energies.  This HBT analysis technique allows one to follow the
trajectories of the test particles after emission up to the thermal
freeze-out in model calculations. It is suitable for examining the
space-time geometry of evolving sources in detail.  In this study we
will use relativistic hydrodynamics with an equation of state of
first-order phase transition to describe the evolution of the
particle-emitting sources produced at GSI-FAIR energies.  We will
use the HBT analysis technique with quantum transport of the
interfering pair to examine the HBT radii and lifetimes of the
sources for different initial energy densities.  Because the central
heavy ion collisions at GSI-FAIR energies are almost full stopped,
we assume that the systems are at rest initially in the
center-of-mass frame and with spherical shape for simplicity.

As compared with pion HBT interferometry kaon HBT interferometry may
present more clearly the source space-time geometry at emission
configuration, because kaons can escape easily from the system after
hadronization and therefore seldom be affected by the multiple
scattering and particle decay.  By comparing the results extracted
from the two-pion and two-kaon HBT interferometry, we find that the
multiple scattering and particle decay lead to a larger radius of
the pion source than that of kaon's.  However, both the HBT
lifetimes of the pion and kaon sources have significant increase
when the initial energy density approaches to the boundary between
the QGP and mixed phase.

The paper is organized as follows.  In section II we describe the
model of equation of state (EOS) used in our calculations.  The
adiabatic paths for the EOS and the space-time configuration of the
evolving system are also discussed in this section.  In section III
we give a brief description to the quantum probability amplitudes in
a path-integral formalism of HBT and the two-pion and two-kaon HBT
results for the evolving sources produced in the collisions at
GSI-FAIR energies.  Finally, the summary and discussion are
presented in section IV.

\section{Equation of State and System Evolution}

For non-dissipative ideal fluid hydrodynamics is defined by the
local conservations of energy-momentum and other conserved
quantities (e.g. entropy, baryon number, and strangeness)
\cite{Ris98,Kol03}.  To solve the hydrodynamical equations one needs
the EOS which gives the relation among the thermodynamical
quantities in the conserved equations \cite{Ris98,Kol03}.  In what
follows we discuss our EOS model and system evolution described by
hydrodynamics and the EOS.

\subsection{EOS model}

In our EOS model the QGP phase is described by an ideal gas of
gluons, $u$, $d$, $s$ quarks and antiquarks, with the constant
vacuum energy $B$ associated with QCD confinement
\cite{Zha00,Ton03}.  The pressure, energy density, and the conserved
charge density in the QGP phase are given by
\begin{equation}
\label{eq-q-eosp}
p^Q = \sum_i p_i(T,\mu_i) - B \,,
\end{equation}
\begin{equation}
\label{eq-q-eose}
\varepsilon^Q = \sum_i \varepsilon_i(T,\mu_i) + B
\,,
\end{equation}
\begin{equation}
\label{eq-q-eosn}
n_A^Q =  \sum_i A_i \  n_i(T,\mu_i) \,,
\end{equation}
where $p_i(T,\mu_i)$, $\varepsilon_i(T,\mu_i)$, and $n_i(T,\mu_i)$
are the pressure, energy density, and number density of particle
species $i$ in the ideal gas with temperature $T$ and chemical
potential $\{\mu_i\}$, $A_i$ is the conserved charge number of the
particle species $i$.  In our calculations we use the quark masses
$m_u=m_d=5$ MeV, $m_s=150$ MeV and the bag constant
$B=(235~\mbox{MeV})^4$ \cite{Ton03}.

For the hadronic phase we adopt the excluded volume model
\cite{Ris91,Hun98,Ton03} and consider the particles $\pi$, $K$, $N$,
$\Lambda$, $\Sigma$, $\Delta$, and their antiparticles.  The
pressure, energy density, and the conserved charge density in the
hadronic phase are given by \cite{Ris91,Hun98,Ton03}
\begin{eqnarray}
\label{eq-h-eosp} p^H = \sum_i p_i(T,\tilde\mu_i)\,,
\end{eqnarray}
\begin{eqnarray}
\label{eq-h-eose} \varepsilon^H = {
\sum_i\varepsilon_i(T,\tilde\mu_i) \over 1 + V_0\sum_i
{n_i(T,\tilde\mu_i) } } \,,
\end{eqnarray}
\begin{eqnarray}
\label{eq-h-eosn} n_A^H = {\sum_i A_i\, n_i(T,\tilde\mu_i) \over 1 +
V_0\sum_i {n_i(T,\tilde\mu_i) } } \,,
\end{eqnarray}
where
\begin{eqnarray}
\tilde\mu_i = \mu_i - V_0\, p^H\,,
\end{eqnarray}
$V_0=(1/2)(4\pi/3)(2a)^3$ is the excluded volume which is assumed to
be the same for all hadrons with $a=0.5$ fm \cite{Ton03}.

For the first-order phase transition, there are Gibbs relationships
in the mixed phase of the QGP and hadron gas.  We have $T^Q=T^H$,
$\mu_{N,\Delta}=3\mu_u$, $\mu_{\Lambda,\Sigma} =2\mu_u+\mu_s$,
$\mu_{\pi^+,\pi^0,\pi^-}=0$, $\mu_{K^+,K^0}=\mu_u-\mu_s$, ..., and
\begin{eqnarray}
\label{eq-m-eosp} p^M=p^Q(T,\mu_u,\mu_s)=p^H(T,\mu_u,\mu_s) \,,
\end{eqnarray}
\begin{eqnarray}
\label{eq-m-eose}
\varepsilon^M = \alpha \,
\varepsilon^Q(T,\mu_u,\mu_s ) + (1-\alpha ) \,
\varepsilon^H(T,\mu_u,\mu_s ) \,,
\end{eqnarray}
\begin{eqnarray}
\label{eq-m-eosn}
n_A^M = \alpha \, n_A^Q(T,\mu_u,\mu_s ) +
(1-\alpha ) \, n_A^H(T,\mu_u,\mu_s ) \,,
\end{eqnarray}
where $\mu_u$ and $\mu_s$ are the chemical potential of $u$ and $s$
quarks, and $\alpha = V_Q / V$ is the fraction of the volume
occupied by the plasma phase.  The boundaries of the coexistence
region are found by putting $\alpha = 0$ (the hadron phase boundary)
and $\alpha = 1$ (the plasma boundary).

Using the thermodynamical relations of ideal gas one can get other
thermodynamical quantities, such as entropy density $s$, in the QGP,
hadronic, and mixed phases from Eqs. [(\ref{eq-q-eosp}) --
(\ref{eq-q-eosn})], [(\ref{eq-h-eosp}) -- (\ref{eq-h-eosn})], and
[(\ref{eq-m-eosp})--(\ref{eq-m-eosn})], and get numerically the EOS
for solving the hydrodynamical equations.

\subsection{Adiabatic paths}

\begin{figure}
\includegraphics[angle=0,scale=0.50]{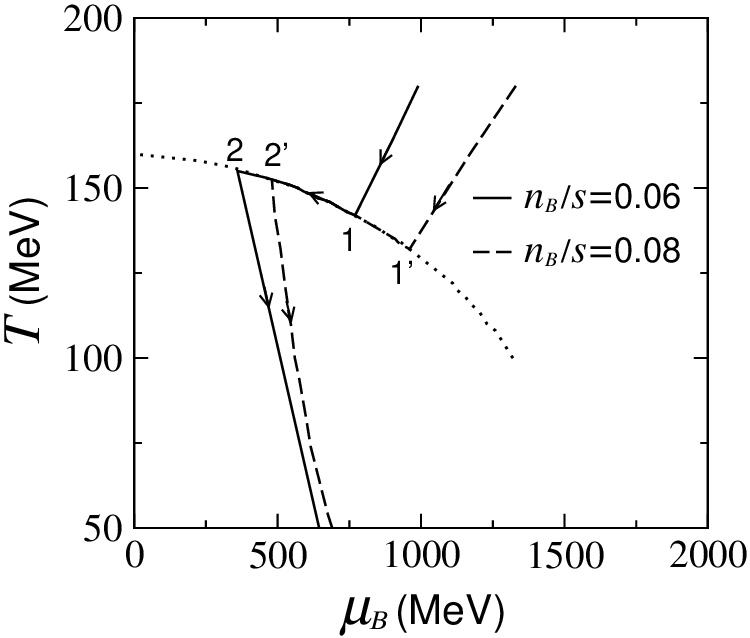}
\caption{\label{fig:adiabat} Adiabatic paths for $n_B/s=0.06$ and
$n_B/s=0.08$.}
\end{figure}

In our model calculations the system evolves from a thermalized
initial state to final freeze-out.  In the absence of dissipation,
the entropy of the system is conserved during evolution.  On the
other hand, the baryon number is also conserved.  So the ratio of
their local densities $n_B/s$ is a constant.  For our EOS model we
show in Fig. 1 the adiabatic cooling paths of the systems with
$n_{B}/s=$0.08 and 0.06, which correspond to the incident energies
about 10 and 30 $A$GeV, respectively \cite{Hun98,Iva06}.  The dotted
line in Fig. \ref{fig:adiabat} is the transition curve between the
QGP and hadron gas.  The mixed phase is on the transition curve from
the endpoint of the QGP branch (point 1 or 1$'$) up to the beginning
of the hadronic branch (point 2 or 2$'$).  The non-trivial zigzag
shape of the trajectories indicates that the system has a re-heating
in the mixed phase \cite{Sub86,Hun98}.  The reason is that at a
certain point ($T,\mu$) on the phase-transition curve, the number of
degrees of freedom and hence the specific entropy in the plasma
phase are larger than the corresponding values in the hadronic
phase.  Hence the temperature must increase during hadronization to
conserve both the total entropy and baryon number simultaneously
\cite{Sub86}.

\begin{figure}
\includegraphics[angle=0,scale=0.50]{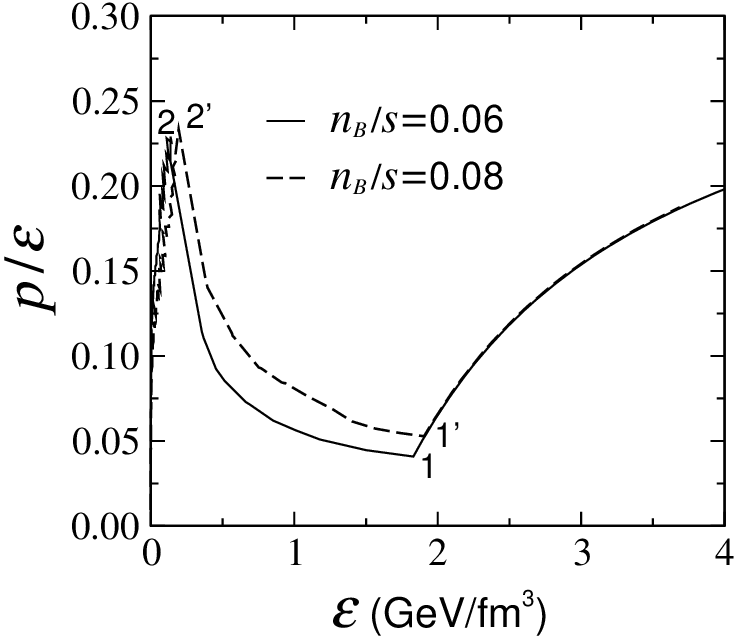}
\caption{\label{fig:pe} The ratio of pressure to energy density
$p/\varepsilon$ for $n_B/s=0.06$ and $n_B/s=0.08$.}
\end{figure}

In Fig. \ref{fig:pe} we show the hydrodynamical relevant relation,
$p/\varepsilon$, for $n_B/s=$0.06 and 0.08.  At the boundaries
between the hadronic and mixed phases (2 and 2$'$) the ratio
$p/\varepsilon$ has maximums.  The minimums of the ratio, named the
``softest points", are at the boundaries between the mixed phase and
QGP (1 and 1$'$), corresponding to $\varepsilon=\varepsilon^{\rm
MQ}=$1.83 and 1.90 GeV/fm$^3$ for $n_{B}/s=0.06$ and 0.08,
respectively.

\subsection{System evolution}

After knowing system EOS we can obtain the solutions of
hydrodynamical equations for the certain initial conditions
\cite{Ris96,Zha04c,Zha04,Efa05}, by using the HLLE scheme
\cite{Sch93,Ris95} and Sod's operator splitting method \cite{Sod77}.
Because the central heavy ion collisions at GSI-FAIR energies are
almost full stopped, we assume that the initial system is at rest in
a sphere with a constant energy density $\varepsilon^0$.  For
$n_{B}/s=0.06$, the incident energy is about 30 $A$GeV
\cite{Hun98,Iva06}.  We investigate the system evolution with the
initial energy densities $\varepsilon^0=4.12\,{\rm
GeV/fm}^3\,>\varepsilon^{\rm MQ}$ and
$\varepsilon^0=\varepsilon^{\rm MQ}=1.83$ GeV/fm$^3$.  The
corresponding initial temperatures are $T^0=$180 and 142 MeV.
Meantime, the corresponding initial baryon chemical potentials are
$\mu_B^0=3\mu_u^0=$990 and 780 MeV.  For the system with
$n_{B}/s=0.08$, the corresponding incident energy about 10 $A$GeV
enable only the initial energy density to approach the region of the
QGP boundary \cite{Hun98,Iva06,Ars07}.  In this case we calculate
the system evolution with the initial energy densities
$\varepsilon^0=\varepsilon^{\rm MQ}=1.90$ GeV/fm$^3$ and
$\varepsilon^0=\varepsilon^{\rm HM}=172$ MeV/fm$^3$, where
$\varepsilon^{\rm HM}$ is the energy density at the boundary between
the hadronic and mixed phases.  The corresponding initial
temperatures are 132 and 152 MeV.  The corresponding initial baryon
chemical potential are 960 and 480 MeV.

\begin{figure}
\includegraphics[angle=0,scale=0.50]{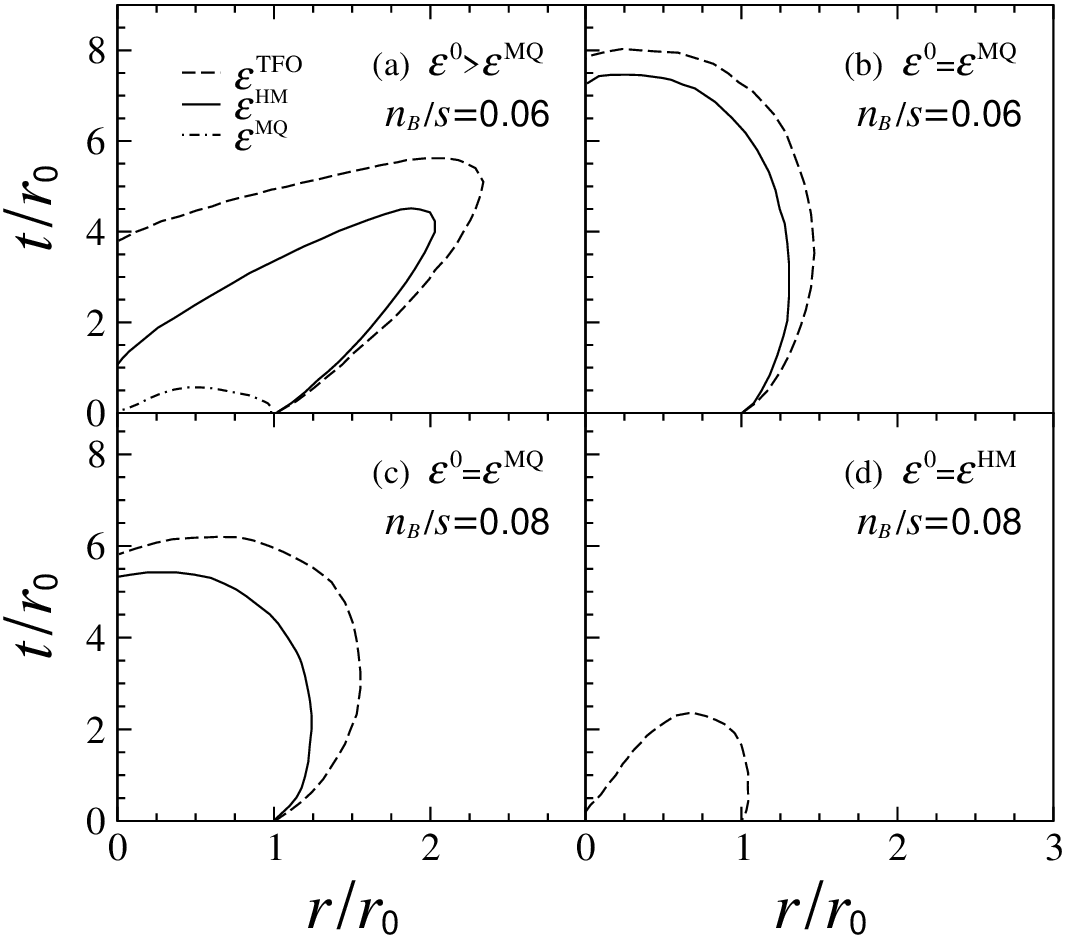}
\caption{\label{fig:evol} The space-time contours of energy density
for the systems with $n_{B}/s=0.06$ (a,b) and $n_{B}/s=0.08$ (c,d),
for the initial energy density $\varepsilon^0 > \varepsilon^{\rm
MQ}$, $\varepsilon^0 = \varepsilon^{\rm MQ}$, and $\varepsilon^0 =
\varepsilon^{\rm HM}$. }
\end{figure}

Figure \ref{fig:evol}(a) and (b) show the space-time contours of the
local energy densities at $\varepsilon^{\rm MQ}$, $\varepsilon^{\rm
HM}$, and $\varepsilon^{\rm TFO}$ for
$\varepsilon^0>\varepsilon^{\rm MQ}$ and
$\varepsilon^0=\varepsilon^{\rm MQ}$ for the system with
$n_B/s=0.06$, respectively.  Here $r_0$ and $\varepsilon^{\rm TFO}$
are the initial system radius and the energy density at the thermal
freeze-out.  One can see that the system takes more time evolving
through the mixed phase (from $\varepsilon^{\rm MQ}$ to
$\varepsilon^{\rm HM}$) than that through the pure QGP phase (from
$\varepsilon^0$ to $\varepsilon^{\rm MQ}$) or the pure hadronic
phase (from $\varepsilon^{\rm HM}$ to $\varepsilon^{\rm TFO}$).  The
duration of the evolution through the mixed phase is larger when the
initial energy density is at the soft point.  Figure
\ref{fig:evol}(c) and (d) show the contours at $\varepsilon^{\rm
HM}$ and $\varepsilon^{\rm TFO}$ for $\varepsilon^0=\varepsilon^{\rm
MQ}$ and $\varepsilon^0=\varepsilon^{\rm HM}$ for the system with
$n_B/s=0.08$, respectively.  For $\varepsilon^0=\varepsilon^{\rm
MQ}$, because the soft point for $n_B/s=0.08$ is higher than that
for $n_B/s=0.06$ (see Fig. \ref{fig:pe}), the evolving time through
the mixed phase for $n_B/s=0.08$ is short than that for
$n_B/s=0.06$.  For $\varepsilon^0=\varepsilon^{\rm HM}$ there is
only hadronic phase.  The evolution is fast because of the larger
$p/\varepsilon$ in the hadronic gas.  In our calculations the energy
density at the thermal freeze-out is taken to be 45 MeV/fm$^3$
\cite{Cle99}, which corresponds to the thermal freeze-out
temperatures 110 and 100 MeV for $n_B/s=$0.06 and 0.08. The initial
system radii are taken to be 6 fm.

\section{HBT interferometry with quantum transport of the interfering pair}

The two-particle Bose-Einstein correlation function $C(k_1,k_2)$ is
defined as the ratio of the two-particle momentum distribution
$P(k_1,k_2)$ to the the product of the single-particle momentum
distribution $P(k_1) P(k_2)$. For an evolving source, using quantum
probability amplitudes in a path-integral formalism,
$P(k_i)~(i=1,2)$ and $P(k_1,k_2)$ can be expressed as
\cite{Won03,Won04,Won05,Zha04c,Yul08}
\begin{eqnarray}
\label{pk1} P(k_i)=\int\!d^4x \,\rho(x)e^{-2\,{\cal I}{m}\,{\bar
\phi}_s(x)}A^2(\kappa x)\,,
\end{eqnarray}
\begin{eqnarray}
\label{pk12} P(k_1,k_2) &=&\int d^4x_1 d^4x_2 \ e^{-2\,{\cal I}{m}\,
{\bar \phi}_s(x_1)} e^{-2\,{\cal I}{m}\,{\bar
\phi}_s(x_2)}\nonumber\\
&&\times\rho(x_1) \rho(x_2)|\Phi(x_1 x_2; k_1 k_2)|^2 ,
\end{eqnarray}
where $\rho(x)$ is the four-dimension density of the
particle-emitting source, $A(\kappa x)$ is the magnitude of the
amplitude for producing a particle with momentum $\kappa$ at $x$,
$e^{-2\,{\cal I}{m}\,{\bar \phi}_s(x)}$ is the absorption factor due
to multiple scattering, and $\Phi(x_1 x_2; k_1 k_2)$ is the wave
function for the two identical boses produced at $x_1$ and $x_2$
with momenta $\kappa_1$ and $\kappa_2$, and detected at $x_{d1}$ or
$x_{d2}$ with momenta $k_1$ and $k_2$, respectively.

In our HBT calculations final identical kaons are considered to be
emitted thermally from the space-time hypersurface at
$\varepsilon^{\rm HM}$ and keep freeze-out after emission. However,
final identical pions (for example $\pi^+$) include the primary
pions emitted from the hypersuface at $\varepsilon^{\rm HM}$ and the
secondary pions from the ``excited-state" particle decays during the
system evolving in hadronic phase until to the thermal freeze-out.
The four-dimension density of the pion source can be expressed as
\cite{Yul08}
\begin{equation}
\rho(x)=n_{\pi}(x)\delta(t-\tau^0)+\sum_{j\ne \pi}D_{j\rightarrow
\pi} n_j(x)\,,
\end{equation}
where $n_i(x)$ and $\tau^0$ are the particle number density and the
hadronization time in local frame, $D_{j\rightarrow \pi}$ is the
product of the decay rate in time and the fraction of the decay
${\tilde d}_{j\rightarrow \pi}$. For example, $D_{\Delta \rightarrow
\pi}= \Gamma_{\Delta}\times \frac{1}{3}$ and $D_{\pi^0\pi^0
\rightarrow \pi^+\pi^-}=v_r n_{\pi} \sigma(\pi^0\pi^0 \rightarrow
\pi^+\pi^-) \times 1$, where $v_r$ is the relative velocity of the
two colliding pions and the cross section $\sigma(\pi^0\pi^0
\rightarrow \pi^+\pi^-)$ is equal to the absorption cross section of
$\pi^+\pi^- \rightarrow \pi^0\pi^0$ \cite{Yul08}.  In calculations
we neglect the contributions of $\Lambda$ and $\Sigma$ decays to the
final pions, because most of them exist until the thermal
freeze-out.

When a test pion propagating in the source it will subject to
multiple scattering with the medium particles in the source.  Based
on Glauber multiple scattering theory \cite{Gla59}, the absorption
factor due to the multiple scattering in Eqs. (\ref{pk1}) and
(\ref{pk12}) can be written as \cite{Won03,Won04,Won05,Zha04c,Yul08}
\begin{eqnarray}
\label{absf} e^{-2\,{\cal I}{m}\,{\bar
\phi}_s(x)}=\exp\Bigg[\!-\!\int_{x}^{x_f}\!
\Big({\sum_i}'\sigma_{\rm abs}(\pi i)\ n_i(x')\Big) d\ell(x')\Bigg],
\end{eqnarray}
where $\sum'_i$ means the summation for all medium particles except
for the test pion along the propagating path $d\ell(x')$,
$\sigma_{\rm abs}(\pi i)$ is the absorption cross section of the
test pion with the particle species $i$ in the medium, and $x_f$ is
the freeze-out coordinate.  In calculations we only consider the
dominant absorption processes for the identical pions, for example
the reactions of $\pi^+\pi^- \to \pi^0 \pi^0$ and $\pi^+ N
\rightarrow \Delta$ for $\pi^+$, as we did in Ref. \cite{Yul08}.

In HBT analyses the variables usually used are the Pratt-Bertsch
variables, $q_{\rm out}$, $q_{\rm side}$, and $q_{\rm long}$
\cite{Pra86,Ber88}.  Here $q_{\rm out}$ and $q_{\rm side}$ are the
components of the relative momentum of identical particle pair in
the directions parallel and perpendicular to the total transverse
momentum of the pair, and $q_{\rm long}$ is the relative momentum
along the beam direction of collision.  Using the Pratt-Bertsch
variables the HBT radius in the side-direction, $R_{\rm side}$,
reflects the transverse size of the source.  However, the HBT radius
in the out-direction, $R_{\rm out}$, is related to not only the
source size, but also the source expanding velocity and lifetime
\cite{Pra86,Ber88}.  So a detailed joint analysis of $R_{\rm out}$
and $R_{\rm side}$ as a function of transverse momentum of the pair
may also provides the information of source dynamics
\cite{Pra86,Ber88,Wie99,Wei00}.  Motivated by investigating the
source lifetime directly and clearly, we use the variables $q=|{\bf
k_1}-{\bf k_2}|$ and $q_0=|E_1-E_2|$ and the simple Gaussian fitting
formula
\begin{eqnarray}
C(q,q_0)=1+{\lambda}\,e^{-q^2R^2-q^2_0\tau^2} \,,
\end{eqnarray}
where $R$, $\tau$, and $\lambda$ are the source HBT radius,
lifetime, and chaotic parameter.

\begin{figure}
\includegraphics[angle=0,scale=0.50]{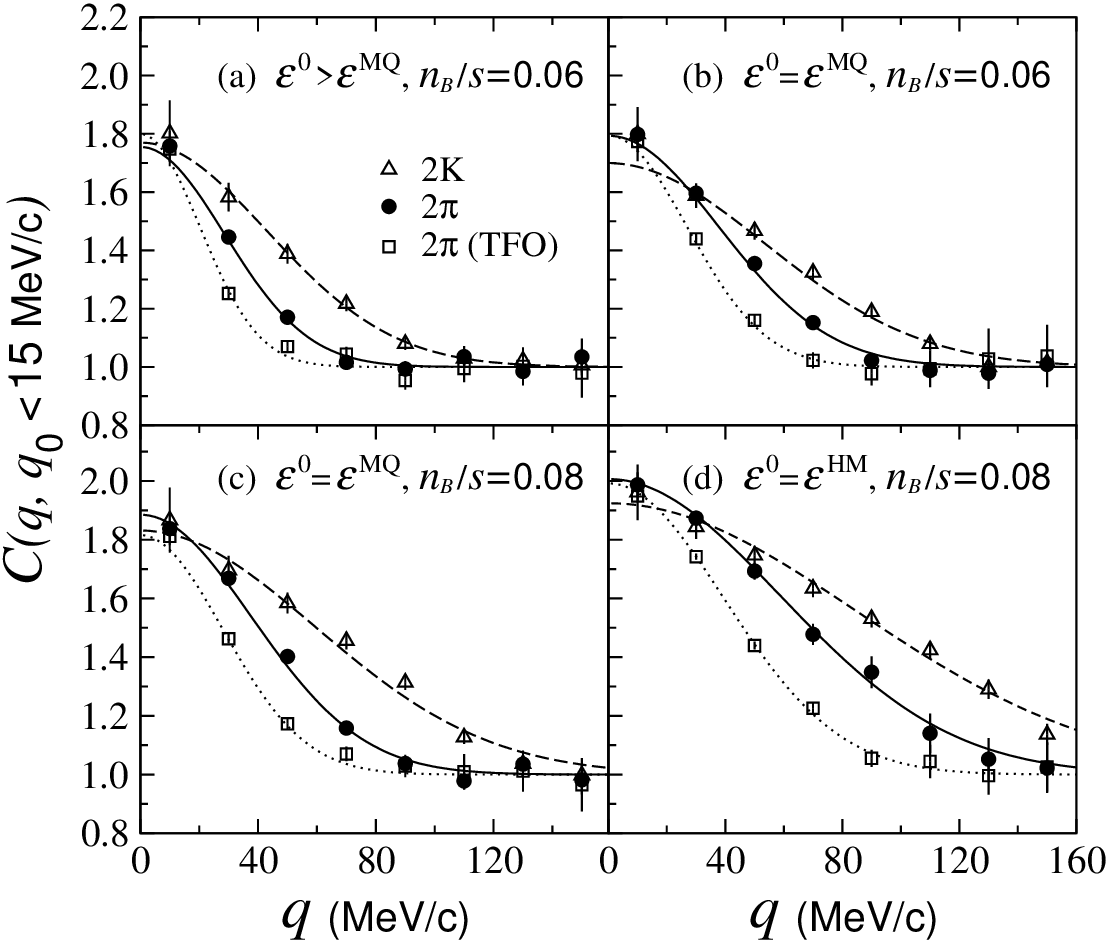}
\caption{\label{fig:cf} The two-particle correlation functions of
kaons and pions for the evolving systems as shown in Fig.
\ref{fig:evol}.}
\end{figure}

From Eqs. (\ref{pk1}) -- (\ref{absf}) we can construct numerically
the two-particle HBT correlation function for each $(q,q_0)$ bin
\cite{Zha04c,Zha04,Zha05,Zha06,Yul08}.  Figure \ref{fig:cf} shows
the two-particle correlation functions $C(\,q,\,q_0<\!15\,{\rm
MeV/c})$ for the evolving sources with $n_B/s=$0.06 and 0.08, and
for $\varepsilon^0 > \varepsilon^{\rm MQ}$, $\varepsilon^0 =
\varepsilon^{\rm MQ}$, and $\varepsilon^0 = \varepsilon^{\rm HM}$.
The symbols {\raisebox{-1.3mm} {\textsuperscript{$\triangle$}}} and
$\bullet$ are the two-kaon and two-pion correlation function
results.  For comparison, the symbols {\raisebox{-1.3mm}
{\textsuperscript{$\Box$}}} present the two-pion correlation
functions calculated with the pions emitted from the thermal
freeze-out (TFO) configuration.  Table \ref{tab:hbt} gives our HBT
fitted results.  One can see that for the same initial conditions,
the two-kaon HBT radius is smaller than that of the two-pion's.  It
is because that the kaons are emitted earlier and from smaller
system configuration.  The multiple scattering and particle decays
during the source evolving in hadronic phase increase the HBT radii
of the pion sources.  Because the pion sources have largest
configuration at TFO, the corresponding HBT radius is the largest
for the certain initial energy density.

\begin{table*}
\caption{\label{tab:hbt}The HBT fitted results.}
\vspace*{0.2cm}
\begin{ruledtabular}
\begin{tabular}{lccc}
{}&$2K$&2$\pi$&2$\pi$(TFO)\\
\hline
\verb"(a)"&$R=3.25\pm0.12$ fm&$R=4.90\pm0.20$ fm&$R=6.95\pm0.30$ fm\\
$\varepsilon^{0}>\varepsilon^{\rm MQ}$&$\tau=5.35\pm0.36$
fm/c&$\tau=7.31\pm0.54$ fm/c&$\tau=7.84\pm0.74$ fm/c\\\vspace{5mm}
$n_{B}/s=0.06$&$\lambda=0.81\pm0.05$&$\lambda=0.83\pm0.05$&$\lambda=0.90\pm0.06$\\
\verb"(b)"&$R=2.61\pm0.09$ fm&$R=3.70\pm0.14$ fm&$R=5.20\pm0.19$ fm\\
$\varepsilon^{0}=\varepsilon^{\rm MQ}$&$\tau=12.50\pm0.68$
fm/c&$\tau=12.53\pm0.58$ fm/c&$\tau=12.43\pm0.65$ fm/c\\\vspace{5mm}
$n_{B}/s=0.06$&$\lambda=0.91\pm0.05$&$\lambda=1.04\pm0.05$&$\lambda=1.03\pm0.05$\\
\verb"(c)"&$R=2.35\pm0.07$ fm&$R=3.70\pm0.10$ fm&$R=4.98\pm0.16$ fm\\
$\varepsilon^{0}=\varepsilon^{\rm MQ}$&$\tau=9.58\pm0.36$
fm/c&$\tau=9.52\pm0.31$ fm/c&$\tau=9.13\pm0.35$ fm/c\\\vspace{5mm}
$n_{B}/s=0.08$&$\lambda=0.98\pm0.05$&$\lambda=1.04\pm0.04$&$\lambda=0.95\pm0.04$\\
\verb"(d)"&$R=1.66\pm0.03$ fm&$R=2.38\pm0.06$ fm&$R=3.54\pm0.04$ fm\\
$\varepsilon^{0}=\varepsilon^{\rm HM}$&$\tau=2.33\pm0.08$
 fm/c&$\tau=2.43\pm0.10$ fm/c&$\tau=1.73\pm0.14$ fm/c\\
$n_{B}/s=0.08$&$\lambda=0.93\pm0.01$&$\lambda=1.02\pm0.02$&$\lambda=1.00\pm0.01$\\
\end{tabular}
\end{ruledtabular}
\end{table*}

For the system with $n_{B}/s=0.06$ we can see that for the initial
energy density $\varepsilon^0 > \varepsilon^{\rm MQ}$ the two-pion
HBT lifetimes are larger than that of two-kaon's.  However, when
$\varepsilon^0$ drops to the soft point $\varepsilon^{\rm MQ}$ the
HBT lifetimes for the pions and kaons increase significantly and
almost are the same, while the corresponding HBT radii decrease. The
reasons are related to their different source space-time geometries
[see Fig. \ref{fig:evol} (a) and (b)] and expanding velocities for
the two kinds of initial conditions.  When the initial energy
density is at the soft point, the source has the smallest expansion
and correspondingly the smallest average spatial size and largest
evolution time from the initial state to the hadronization.  Because
there is no the influence of the system evolution on kaons after
hadronization, the increase of the HBT lifetime of the kaons
reflects the prolongation of the system evolution in the mixed
phase.  In Table I, the results of chaotic parameter $\lambda$ for
$\varepsilon^0 > \varepsilon^{\rm MQ}$ are obviously smaller than
unit.  This is mainly because that the particle-emitting sources in
this case are much different from Gaussian distribution.

For the system with $n_{B}/s=0.08$, the two-pion and two-kaon HBT
lifetimes for $\varepsilon^0=\varepsilon^{\rm MQ}$ are much larger
than those for $\varepsilon^0=\varepsilon^{\rm HM}$.  Also, the HBT
radii for $\varepsilon^0=\varepsilon^{\rm MQ}$ are larger than the
corresponding results for $\varepsilon^0=\varepsilon^{\rm HM}$.  The
main reason for these results is that the space-time configuration
for $\varepsilon^0=\varepsilon^{\rm HM}$ is much smaller than that
for $\varepsilon^0=\varepsilon^{\rm MQ}$ [see Fig. \ref{fig:evol}
(c) and (d)].  Our HBT investigations indicate that both for
$n_B/s=$0.06 and 0.08, the maximums of the HBT lifetimes appear when
the initial energy density reaching at the soft points. Because the
ratio $p/\varepsilon$ at the soft point $\varepsilon^{\rm MQ}$ for
$n_B/s=0.08$ is higher than that for $n_B/s=0.06$ (see Fig.
\ref{fig:pe}), the maximal HBT lifetimes for $n_B/s=0.08$ are
smaller than those corresponding lifetimes for $n_B/s=0.06$.

For the systems with the $n_B/s$ values between 0.06 and 0.08, we
also find that their HBT lifetimes increase significantly when
$\varepsilon^0$ approaches at the corresponding soft points (between
1 and 1$'$ in figure \ref{fig:pe}).  The maximums of the HBT
lifetime are about 10 fm/c much larger than the results for
$\varepsilon^0 > \varepsilon^{\rm MQ}$ and $\varepsilon^0 =
\varepsilon^{\rm HM}$ [\ in Table I (a) and (d)]. Based on the
evolving trajectories calculated by hydrodynamics \cite{Iva06}, the
events with the initial thermalized states staying in the mixed
phase of the QGP and hadronic gas will happen when the incident
energies are between 10 and 30$A$ GeV (see the figure 19 of
\cite{Iva06}, the beginnings of the bold parts of the trajectories
for the 10 and 30$A$ GeV are at the two sides of the transition
region, respectively).  Correspondingly, the ratio of $n_B/s$ is
about between 0.08 and 0.06 (see the figure 18 of \cite{Iva06}).
From our model calculations, the maximums of the two-pion and
two-kaon HBT lifetimes will be observed simultaneously in GSI-FAIR
energy range when the initial energy density is tuned to the soft
point.

\section{Summary and discussion}

Recently the heavy ion collisions at the energies between AGS and
the top-energy SPS attract special attention, for example the SPS
and RHIC low energy programs \cite{Mic06,San06,Ste06,Ste08} and the
project of GSI-FAIR \cite{Hoh05,Pet06,Fri06,Ars07,Ros07,Hen08}.  In
this energy range it is expected that the heavy ion collisions may
produce the QGP with high baryon density, which is differen from
that have been observed in RHIC and top-energy SPS experiments.
Correspondingly, the phase transition from the high-baryon-density
QGP to hadronic matter is the first-order transition, which will
lead to a system evolution much differen from the crossover
transition happened in the low baryon density region at RHIC and top
SPS energies.

Using the technique of quantum transport of the interfering pair we
examine the two-pion and two-kaon HBT interferometry for the
particle-emitting sources produced in the heavy ion collisions at
GSI-FAIR energies.  We use relativistic hydrodynamics with the EOS
of first-order phase transition between the QGP and hadronic gas to
describe the system evolution.  The two-particle HBT correlation
functions are calculated with the quantum probability amplitudes in
a path-integral formalism, where the effects of particle decay and
multiple scattering are taken into consideration.  We find that both
the HBT radii of pions and kaons increase with the system initial
energy density.  The particle decay and multiple scattering lead to
the larger HBT radii of pions than the corresponding HBT radii of
kaons.  The HBT lifetimes of the pion and kaon sources are sensitive
to the initial energy density.  They are significantly prolonged
when the initial energy density is tuned to the soft point.  Our
model calculations indicate that this significant prolongation of
the HBT lifetimes of pions and kaons will be observed in the heavy
ion collisions at GSI-FAIR energies.

As a useful space-time probe HBT interferometry has been extensively
used in high energy heavy ion collisions.  However, there are still
some open problems on HBT analysis technique and the understanding
of HBT results.  The HBT measurements at RHIC indicate that the
values of the ratio of the transverse HBT radii $R_{\rm out}$ to
$R_{\rm side}$ are smaller than those from the hydrodynamical
calculations \cite{STA01a,PHE02a,PHE04a,STA05a}.  Various models and
techniques have been put forth to explain the RHIC HBT puzzle
\cite{Sof02J,Hei02,Lin02,Tea03,Cso03,Mol04,Soc04,Zha04,Kap04,Gra05,Pra05,Zha06,Fro06,Liq07,Bro08}.
At GSI-FAIR energies, the heavy ion collisions are almost full
stopped. We used an approximation of spherical evolving sources and
assumed that the initial states are static and uniform in our
calculations. It would be interesting to consider more reasonable
evolving sources and study the effect of initial conditions on the
HBT results in future investigations. Also, a systematical
investigation to HBT interferometry in different energy ranges will
be of great interest.

\begin{acknowledgments}
The authors would like to thank Dr. C. Y. Wong for helpful
discussions.  This research was supported by the National Natural
Science Foundation of China under grants 10575024 and 10775024.
\end{acknowledgments}

\end{document}